\documentclass[12pt,prb,aps]{revtex4}
\usepackage{amsmath}

\usepackage[pctex32]{graphics, graphicx}

\begin{document}

\title{Arago (1810): the first experimental result against the ether}

\author{Rafael Ferraro}
\altaffiliation{Member of Carrera del Investigador Cient\'{\i}fico
(CONICET, Argentina)} \email[Electronic mail:
]{ferraro@iafe.uba.ar} \affiliation{Instituto de Astronom\'\i a y
F\'\i sica del Espacio,
Casilla de Correo 67, Sucursal 28, 1428 Buenos Aires, Argentina\\
and Departamento de F\'\i sica, Facultad de Ciencias Exactas y
Naturales, Universidad de Buenos Aires, Ciudad Universitaria,
Pabell\' on I, 1428 Buenos Aires, Argentina}

\vskip1cm

\author{Daniel M. Sforza}
\email[Electronic mail: ]{sforza@ucla.edu} \affiliation{Laboratory
of Neuro Imaging, Department of Neurology, David Geffen School of
Medicine, UCLA, 710 Westwood Plaza, Los Angeles, CA 90095-1769,
USA}

%\date{}

\bigskip

\begin{abstract}
95 years before Special Relativity was born, Arago attempted to
detect the absolute motion of the Earth by measuring the
deflection of starlight passing through a prism fixed to the
Earth. The null result of this experiment gave rise to the
Fresnel's hypothesis of an ether partly dragged by a moving
substance. In the context of Einstein's Relativity, the sole frame
which is privileged in Arago's experiment is the proper frame of
the prism, and the null result only says that Snell's law is valid
in that frame. We revisit the history of this premature first
evidence against the ether theory and calculate the Fresnel's
dragging coefficient by applying the Huygens' construction in the
frame of the prism. We expose the dissimilar treatment received by
the ray and the wave front as an unavoidable consequence of the
classical notions of space and time.
\end{abstract}


\maketitle

\bigskip

\section{Introduction}\label{intro}
Einstein once declared that the Fizeau measurements on the speed
of light in moving water ``were enough'' for experimentally
supporting the relativistic notions of space and
time.\cite{shankland} To understand the meaning of this statement,
it should be remembered that before Einstein's work the {\it
lumiferous ether} was the supposed material substrate where the
luminous perturbation propagates. At first, scientists believed
that the ether was not modified by a moving body, but it pervaded
all the bodies remaining itself immutable. This property, together
with its universality, lent the ether the status of absolute frame
(in the newtonian sense). However, in 1818, Augustin Jean Fresnel
(1788-1827) \cite{fresnel} stated that the ether {\it inside} a
substance is partly dragged when the substance moves with respect
to the (exterior) universal ether. For a transparent substance
moving with ``absolute'' velocity $V$, the ether in the substance
moves with ``absolute'' velocity $(1-n^{-2})\, V$, where $n$ is
the refraction index of the substance. The dragging was
``measured'' in moving water by Fizeau (1851)\cite{fizeau} and
Michelson and Morley (1886). \cite{mm1886,foot1} Special
Relativity abolished the concept of ether in 1905, and these speed
measurements are no longer explained as the result of partial
dragging but as a relativistic composition of velocities. This is
the reason for Einstein's assertion.

The partial dragging of ether is mentioned in the textbooks in
connection with the experiments performed by Hoek (1868)
\cite{hoek} and Airy (1871) \cite{airy}. These experiments
integrate the series of attempts to detect the absolute motion of
the Earth, which culminated with the crucial Michelson-Morley
experiment of 1887 \cite{mm1887} and its repetitions
\cite{swenson}. Differing from Michelson-Morley's array, Hoek and
Airy forced the light to pass by water to succeed in getting a
device which was sensible to the first order in $V/c$. Since Hoek
and Airy did not detect any motion, the pre-relativistic Physics
justified the null result by resorting to the Fresnel partial
dragging of ether. In fact, when the dragging enters into the
scene, the effects of first order in $V/c$ cancel out and the
expected results turn out to be of a larger order. Thus the
absolute motion of the Earth could be undetectable due to the
inaccuracy of the experimental devices. This argument helped to
keep ether theories sustainable for a time.

It would be completely remarkable that a property of the ether
conceived by Fresnel in 1818 could play, 50 years later, a so
important role in supporting the ether theory. However it should
be said that, far from being visionary, Fresnel's hypothesis was
formulated to explain an already existent null result obtained by
Arago in 1810 \cite{arago} (for a historical analysis of Fresnel's
ether dragging theory see Ref. \onlinecite{pedersen}). Dominique
Fran\c{c}ois Jean Arago (1786-1853) was a pioneer of the
experiments that searched for differences in the speed of light.
Arago considered that light coming from different sources would
have different velocities. In the framework of the corpuscular
model, on which Arago based his arguments and experiments, these
differences could be due to diverse reasons: the velocity of the
source could influence the speed of the emitted light (emission
theories), the speed of starlight could be affected by the
gravitational field of the star, etc. Arago remarked that Bradley
aberration\cite{bradley} was not adequate to detect differences of
speed in starlight (a difference of 1/20 would produce a change of
1'' in the aberration angle, which was inaccessible to the
instruments of that time). Instead, Arago noted that such
differences could be registered by measuring the deflection of
starlight by a prism: ``...by remembering that the deviation
experienced by the luminous rays when they obliquely penetrate the
diaphanous bodies is a determined function of their original
velocity, it can be understood that the observation of the total
deviation to which they are subjected when they go through a
prism, provides a natural measure of their
velocities''.\cite{arago2} Arago did not succeed in detecting
different speeds of light, but he soon realized that the method
could be applied to make evident the motion of the Earth:
``...when the refractive bodies are in motion, the refraction
experienced by a ray should not be calculated with its absolute
velocity, but with this same velocity increased or diminished by
the one of the body, i.e. with the relative velocity of the ray.
... I devoted myself, in my experiments, to become evident the
differences that have to result from the orbital motion of the
Earth...''.\cite{arago2} The absolute motion referred to by Arago
in the former sentences is the motion in the newtonian absolute
space; Arago does not mention the ether in his work because it has
no place in the corpuscular model.

Arago's experiment was unable to exhibit the motion of the Earth,
and Fresnel elaborated an hypothesis to explain this null result
in the context of the wave theory of light with an elastic ether.
Given the historic importance of Arago's experiment as the first
attempt to detect the absolute motion of the Earth and the trigger
for the Fresnel hypothesis, the lack of references to Arago's work
in the textbooks is intriguing , where only tangential mentions
can be found. \cite{born} To repair this appreciable oblivion, we
describe Arago's experiment and its results in Section II. Section
III is devoted to revisit the Fresnel hypothesis. In Section IV,
we show how the Fresnel partial dragging of ether can explain the
null result of Arago's experiment. We do not follow here the
demonstration given by Fresnel, who analyzes the deviation in the
frame of the universal ether by computing the time of traveling of
the rays. Instead we propose to study the envelope of the
(anisotropic) secondary wave fronts in the frame of the prism. In
this way, we expose the unacceptable dissimilar treatment received
by the ray and the wave front, otherwise hidden in Fresnel's
original deduction.

\section{Arago's experiment}\label{exp}
As was already mentioned, Arago thought that the refraction of a
light ray by a prism depended on the velocity of the prism.
Specifically, Arago stated that only the velocity of light
relative to the prism should enter the refraction law. Thus, since
the prism moves together with the Earth, a starlight ray would
suffer different deviations depending on the direction of the ray
with respect to the motion of the Earth. These differences would
turn out to be maximal between rays traveling in the same
direction as the Earth and rays traveling in the opposite
direction.

Arago mounted an achromatic prism on the objective lens of a
telescope, and he observed several stars through the prism. By
comparing this apparent position with the real position of each
star, Arago determined how the prism deviated the light rays
coming from each star. He expected different angles of deviation
for different stars. However, Arago declared: ``the rays of all
stars are subjected to the same deviations; the slight differences
found do not follow any law''. \cite{arago2}

\begin{figure}
  \centering
  \includegraphics[scale=0.7]{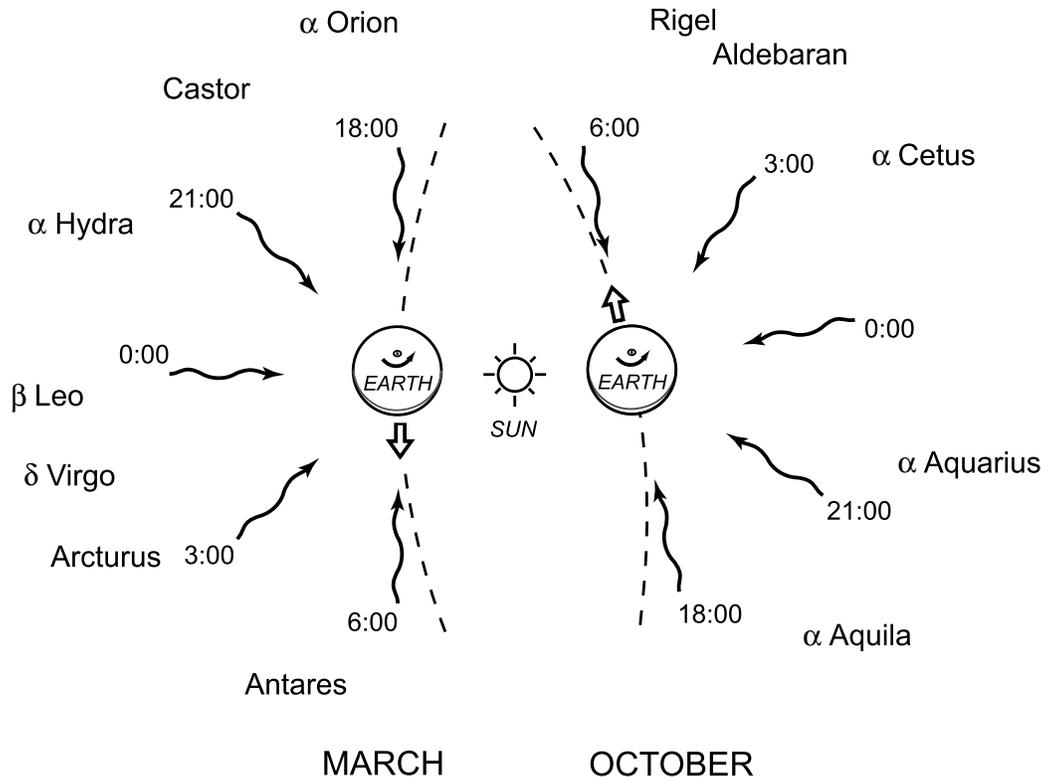}\\
  \caption{This diagram shows the light from different stars and the Earth's
  orbital motion when Arago's measurements were made on March 27 and October 8, 1810. Wavy lines
  represent light from stars (projected on Earth's orbital plane),
  and the hollow arrows correspond to the Earth's orbital velocity.
  The different Galilean compositions of motions were expected to
  result in different starlight deviations by the prism.
  }\label{Fig1}
\end{figure}

Arago made the measurements in two different times of the year. On
March 19 and March 27, he utilized a prism of 24$^\circ$ and a
mural telescope to measure the distances from the stars up to the
zenith. On October 8, Arago used a larger prism to cover a half of
the objective lens of an improved telescope (namely a {\it
repeating circle}), which permitted more precise measurements. In
this way, the telescope could receive the starlight directly from
the star or previously deviated by the prism (the angle between
these two different directions, corrected by the time elapsed
between both measurements, was the deviation angle). The results
are reproduced in Tables \ref{mediciones marzo19}, \ref{mediciones
marzo27} and \ref{mediciones octubre} (the measurements are
displayed in the way Arago did it; besides we added the
approximated times when the stars passed by the Paris meridian to
help in locating the stars in the sky, as Fig. \ref{Fig1} shows).
The motion of the earth around the Sun (30 km/s) was enough to
generate meaningful differences in the deviation angle of stars
passing the meridian at 6.00 a.m. and stars passing it at 6.00
p.m. In the first case, the starlight is opposite to the motion of
the Earth; in the second one, the starlight goes in the same
direction than the Earth. For these two extreme cases, Arago
expected differences of 12'' with the prism of 24$^\circ$ used in
March, and 28'' in the measurements performed in October. However,
there were no traces of any composition of motion entering Snell's
law in the results obtained by Arago.

Arago convinced himself that the only possible interpretation of
his null result, in the context of the corpuscular model, was that
the sources emit light with all sort of velocities. But our eyes
are only sensitive to a narrow band of them. Thus, we are always
detecting the same kind of corpuscles, and no differences can be
found in their refraction. Concerning the wave theory of light,
Arago said that ``the explanation for the refraction in this
theory is based on a simple hypothesis that is very difficult to
submit to calculation. Therefore, I cannot determine in a precise
way whether the velocity of the refractive body has some influence
on the refraction...'' \cite{arago2}

\section{Fresnel's hypothesis}\label{drag}

In 1818, Fresnel published a letter  addressed to Arago,
\cite{fresnel} where he discusses the interaction between the
ether and the bodies. To begin with, Fresnel mentions that Arago's
null result could be easily understood in the context of the wave
theory of light by accepting that the Earth imparts its motion to
the surrounding ether. In that case, the prism would be at rest in
the local ether, and no differences of velocities would appear
(the speed of light is a property of the ether). However, in spite
of its simplicity, this hypothesis would hinder the understanding
of the starlight aberration: ``So far I could not conceive this
phenomenon [the starlight aberration], apart from supposing that
the ether freely passes through the globe, and that the velocity
imparted to this subtle fluid is nothing but a small part of the
one of the Earth, which does not exceed the hundredth for
example.''\cite{fresnel2,foot2} Although the Earth should be
pervaded by an ether flow, Fresnel says that light, which is an
ether vibration, does not propagate inside the Earth due to an
interference of secondary waves. Concerning the transparent media,
Fresnel says: ``it is evident that the placing of water among the
particles, which favors the propagation of luminous vibrations,
must be a little obstacle to the establishment of an ether
flow.''\cite{fresnel2} So, Fresnel thought that the way the ether
flows through a body depends on the properties of the body.

In contrast to the corpuscular model, Snell's law implies for the
wave theory of light that, the bigger the refractive index of a
transparent substance, the slower the light propagates in its
interior.\cite{foot3} Fresnel considered the ether as an elastic
material. It is well known that the velocity of waves propagating
in an elastic material is proportional to $\rho^{-1/2}$, where
$\rho$ is the density of the material (the ether, in our case).
This means that the density of the ether should be bigger in water
or glass than in air. At this point, Fresnel's hypothesis about
the dragging of ether enters the scene: ``only a part of this
medium [the interior ether] is dragged by our globe, those
constituting the excess of its density with respect to the
environmental ether.'' ``... when only a part of the medium moves,
the velocity of propagation of the waves should be increased just
by the velocity of the center of gravity of the
system.''\cite{fresnel2} Since the Earth does not appreciably drag
the ether, there is not excess of ether inside the globe. Instead,
the ether is in excess inside the prism, as revealed by the lower
speed of light in the substance. The excess of ether in the prism
is completely dragged when the prism moves. Only the velocity of
the center of gravity of the system ``environmental ether - excess
of ether'' is imparted to the speed of light in the prism. Let
$\rho^\prime$ be the density of ether inside the prism, and $\rho$
the density of the environmental ether. Thus the excess of ether
is $\rho^\prime-\rho$. If $V$ is the velocity of the prism
relative to the environmental ether (absolute velocity of the
prism), then the speed of the center of gravity of the system
``environmental ether - excess of ether'' is $(\rho^\prime-\rho)\,
V/\rho^\prime$ (the excess of ether is completely dragged). This
motion is imparted to the light traveling inside the prism:
\begin{equation}\label{vdrag}
v_{drag}=\left(1-\frac{\rho}{\rho^\prime}\right)V=\left(1-n^{-2}\right)V,
\end{equation}
the last member being a consequence of the relation between
density of the elastic ether and velocity of propagation of the
waves. The coefficient $1-n^{-2}$, where $n$ is the refractive
index of the transparent substance, is the so called Fresnel
dragging coefficient.

\section{Recovering Snell's law in the frame of the prism}\label{snell}
Independently on the way of conceiving the partial dragging of
ether, the only tangible effect of the dragging is the restoration
of Snell's law in the proper frame of the prism. In this Section,
we will show how the restoration works. Our treatment will differ
from the one followed by Fresnel. Whereas Fresnel studied ray
paths in the frame of the environmental ether (where the prism
moves), we instead will use the Huygens' construction in the
proper frame of the prism. Although the calculation will follow
the precepts of classical Physics, for later discussions we want
to control rays and wave fronts in the sole frame that is regarded
as privileged in Special Relativity. Like Fresnel, we will
concentrate on the simplest case where the incident wave normally
strikes the face of the prism and the velocity of the prism $\bf
V$ is parallel to this incident direction. In this way the first
face does not play any role.

\begin{figure}
  \centering
  \includegraphics[scale=0.6]{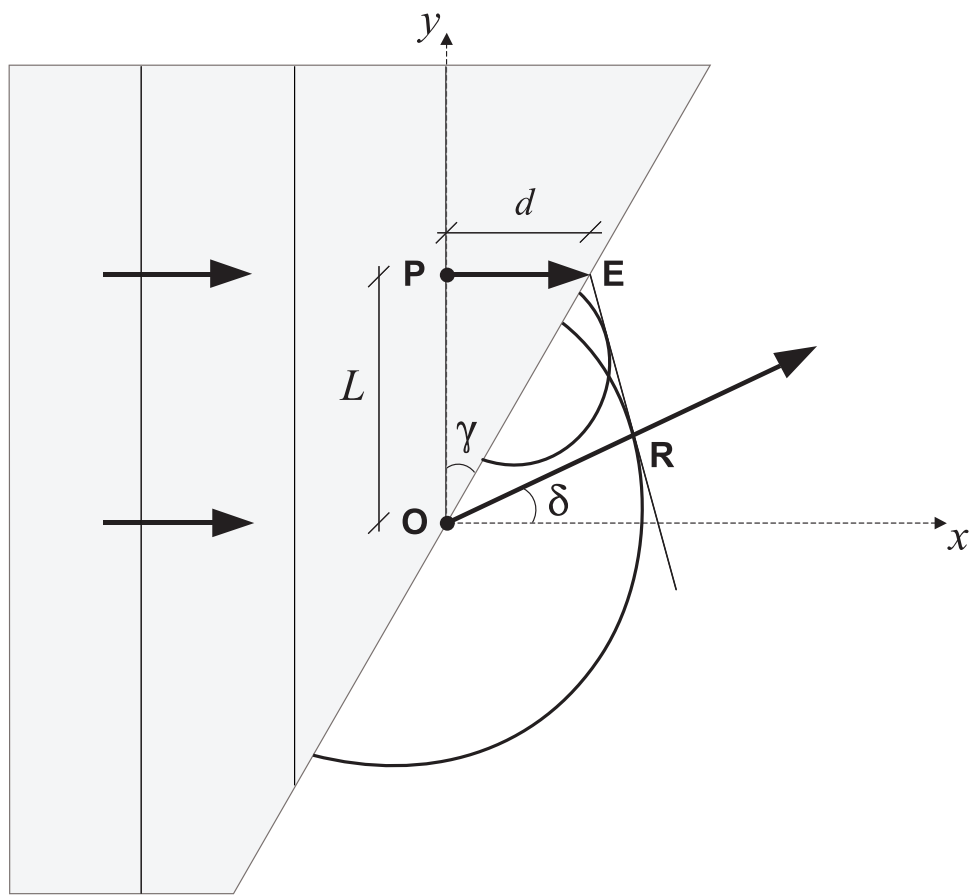}\\
  \caption{Huygens' construction in the frame of the prism.
  }\label{Fig2}
\end{figure}

If the ether were not dragged by the prism, the speed of light
inside the prism would be ${\bf c}/n$ in the frame of the ether
and ${\bf c}/n\, -\, {\bf V}$ in the frame of the prism (Galilean
composition of velocities). However if the prism partly drags the
ether, then the respective velocities change to ${\bf c}/n\, +\,
f\, {\bf V}$ and ${\bf c}/n\, -\, {\bf V}\, +\, f\, {\bf V}$,
where $f\, <\, 1$ is a coefficient that quantifies the magnitude
of the dragging. According to the Huygens' construction, when the
plane wave front traveling inside the prism reaches the second
face, each point of this face becomes a secondary emitter. The
emerging front wave is given by the envelope of the secondary wave
fronts, and the rays go from the secondary emitters to the points
where the envelope touch the secondary wave fronts (see Fig.
\ref{Fig2}). Like Fresnel, we will demand the restoration of
Snell's law at the lowest order in $V/c$. For this, we must choose
$f$ in such a way that the ray direction does not depend on $V$ at
that order. We expect to find $f\, =\, 1\, -\, n^{-2}$.

In order to compute the envelope, we should pay attention to the
fact that secondary wave fronts are not spheres in the frame of
the prism. This is due to the ether wind that generates an
anisotropy in the wave propagation. Concretely, if ${\bf
k}\cdot{\bf r}-\omega\, t$ is the invariant phase of a wave in the
frame of the ether, then, by Galileo transforming the coordinates
to a frame moving with velocity ${\bf V}$, we obtain:
\begin{equation}\label{phase}
{\bf k}\cdot{\bf r}\, -\, \omega\, t\, =\, \vert{\bf
k}\vert\left(\hat{\bf n}\cdot{\bf r}\, -\, c\, t\right)\, =\,
\vert{\bf k}\vert\left(\hat{\bf n}\cdot({\bf r}^\prime\, +\, {\bf
V}t)\, -\, c\, t\right)\, =\, \vert{\bf k}\vert\left(\hat{\bf
n}\cdot{\bf r}^\prime\, -\, (c-\hat{\bf n}\cdot{\bf V})t\right)\,
,
\end{equation}
where $\hat{\bf n}$ is the direction of propagation. So the phase
velocity in a frame moving with velocity {\bf V} relative to the
ether is
\begin{equation}\label{velocity}
v^\prime_{\ phase}\, =\, c\, -\, \hat{\bf n}\cdot{\bf V}\, .
\end{equation}
This means that a wave front emitted by a point-like source at the
coordinate origin is not a sphere in a frame which moves with
(absolute) velocity ${\bf V}$, but is described by the parametric
equations (see Fig. \ref{Fig2})
\begin{eqnarray}\label{front}
x(\delta)&=&\cos\delta\, (c\, -\, V\, \cos\delta)\, t\cr
y(\delta)&=&\sin\delta\, (c\, -\, V\, \cos\delta)\, t
\end{eqnarray}
where ${\bf V}\, =\, V\, \hat{\bf x}$, and $\hat{\bf n}\, =\,
\cos\delta\ \hat{\bf x}\, +\, \sin\delta\ \hat{\bf y}$.

Fig. \ref{Fig2} shows plane wave fronts traveling through the
transparent substance with velocity $c/n\, -\, (1-f)\, V$. When a
plane front arrives at O, this point becomes an emitter of a
secondary wave front. After a lapse $t$, the same plane front
arrives at E; in the meantime, the secondary wave front emitted at
O has evolved according to Eq. (\ref{front}). The envelope ER is
not orthogonal to the ray OR. Actually the emerging rays are
orthogonal to the envelope only in the frame of the ether.
Instead, in a moving frame the rays suffer aberration. The
deviation angle $\delta$ is the value of the parameter of
Eq.(\ref{front}) such that the straight line passing by E touches
one and only one point of each secondary wave front. In general a
straight line passing by E satisfies the equation
\begin{equation}\label{straight}
y\, =\, m\, (x\, -\, d)\, +\, L\, =\, m\, x\, +\, d\,
({\frac{1}{\tan\gamma}}\, -\, m)
\end{equation}
where $m$ is some slope for the straight line. By replacing $d\,
=\, (c/n\, -\, (1-f)\, V)\, t$ and $x$, $y$ by the values given in
Eq. (\ref{front}), one obtains an equation for $\delta(m)$ telling
the values of the parameter $\delta$ for which the straight line
passing by E intersects the secondary wave front emitted from O.
This equation is
\begin{equation}\label{slope}
\sin\delta\, (c\, -\, V\, \cos\delta)\, =\, m\, \cos\delta\, (c\,
-\, V\, \cos\delta)\, +\, (\frac{c}{n}\, -\, (1-f)\, V)\,
({\frac{1}{\tan\gamma}}\, -\, m)
\end{equation}
Since the envelope is tangent to the wave front, we will look for
the maximum value of $m$ which makes sense in the  Eq.
(\ref{slope}). By regarding $m$ as a function of $\delta$, the
solution is $\delta$ such that $dm/d\delta\, = \, 0$. Thus, once
Eq. (\ref{slope}) was derivated with respect to $\delta$, one
replaces $dm/d\delta\, = \, 0$ to get
\begin{equation}\label{m}
c\, \cos\delta\, -\, V\, \cos^2\delta\, +\, V\, \sin^2\delta\, =\,
-m\, c\, \sin\delta\, +\, 2\, m\, V\, \sin\delta\, \cos\delta
\end{equation}
By keeping only terms which are linear in $V/c$, the slope of the
envelope turns out to be
\begin{equation}\label{msolution}
m\, \simeq\, -\frac{1}{\sin\delta}\, (\cos\delta\, +\, \frac{V}{
c})
\end{equation}
where now $\delta$ is the direction of the light ray OR. The so
obtained value of $m$ can be replaced in Eq. (\ref{slope}) to
calculate the direction $\delta$ as a function of the absolute
velocity of the substance $V$ and the dragging coefficient $f$.
Again we will keep only the terms that are linear in $V/c$. After
a bit of algebra, the direction $\delta$ of the emerging light ray
results
\begin{equation}\label{ray}
\frac{1}{\sin\delta}\, (1\, -\, \frac{1}{n}\, \cos\delta)\, -\,
\frac{1}{n\tan\gamma}\, \simeq\, \left[\frac{1}{n\sin\delta}\, -\,
(1-f)\,
\left({\frac{1}{\tan\gamma}}+{\frac{\cos\delta}{\sin\delta}}\right)\right]\,
\frac{V}{c}
\end{equation}
Therefore the direction of the emerging ray is not sensitive to
the first order in $V/c$ --the requirement to explain the null
result of the Arago's experiment-- when
\begin{equation}\label{f}
\frac{1}{n \sin\delta}\, -\, (1-f)\,
\left({\frac{1}{\tan\gamma}}+{\frac{\cos\delta}{\sin\delta}}\right)\,
=\, 0
\end{equation}
In this case, the direction of the emerging ray fulfills
\begin{equation}\label{Snell}
{\frac{1}{\sin\delta}}\, (1\, -\, \frac{1}{n}\, \cos\delta)\, -\,
\frac{1}{n\tan\gamma}\, \simeq\, 0
\end{equation}
i.e., $\sin(\gamma+\delta)\, = \, n\, \sin\gamma$, as it is easily
obtained from Snell's law. From the last two equations, we get the
Fresnel's dragging coefficient $f\, =\, 1\, -\, n^{-2}$.

\section{Conclusions}

Fresnel's success in explaining Arago's null results had very
important consequences. In the first place, the dominant
corpuscular light theory suffered a setback. Arago's null results
could not be reasonably understood in the context of the
corpuscular model. Thus, the less accepted wave theory of light
promoted by Fresnel acquired relevance. The ether dragging theory
was an important step to impose the new framework, which became
even more accepted after Fresnel's mathematical description of
diffraction a few years later (these facts made Arago himself
 to start supporting the wave theory of light). However the ether
theory was not free of problems (for instance, since the
refraction indexes of the substances vary with frequency,
 Fresnel's model would require a different ether for each
frequency, which seems implausible), which were eventually solved
with its abolition by Special Relativity. Second, the predominance
of the wave theory of light put the relationship of ether-absolute
motion on the center of the scene.

In Fresnel's wave theory, there is a privileged reference frame
(the ether) where light propagates with velocity $c$ in any
direction. When a transparent substance is at rest in the ether,
Snell's law of refraction is satisfied in the substance. In the
case where the substance is moving in the environmental ether,
there exists a subtle balance between aberration and partial
dragging of the ether whose net result is to hide the absolute
motion, because Snell's law remains valid in the frame of the
substance in first order in $V/c$. Thus, the ether partial
dragging diminishes the privileged status of the ether as the
universal frame to which all movements can be referred to.

The partial dragging of ether renders valid Snell's law in the
frame of the moving substance in first order in $V/c$, because the
ray deviates the same angle in a substance at rest in the ether.
However an annoying asymmetry between ray and wave front is left,
even in first order in $V/c$: while the emerging ray is
perpendicular to the wave front in the frame of the ether, this is
not true in the frame of the prism, as a consequence of the
anisotropic character of the secondary wave fronts in that frame.
In other words, the envelope does not accomplish Snell's law in
the frame of the prism, which highlights the incompatibility of
Galilean transformations and light theory.

In contrast, Special Relativity states that the only privileged
frame in diffraction phenomena is the frame of the refracting
substance, and that Snell's law holds exactly true in that
reference frame. In addition, the emerging ray (just like any ray
propagating in vacuum) is perpendicular to the wave front in any
inertial reference system thanks to the relativistic notions of
space and time. Thus, the ray and the envelope receive the same
treatment. Remarkably, Special Relativity is comfortable with both
the corpuscular and wave models of light. From the corpuscular
point of view, the aberration of light is a composition of
motions. In the context of the wave model, the aberration of light
is the change of orientation of the wave front associated with the
relativity of the simultaneity of secondary wave emissions.

\begin{acknowledgments}
R.F. was supported by Universidad de Buenos Aires (UBACYT X103)
and Consejo Nacional de Investigaciones Cient\'{\i}ficas y
T\'{e}cnicas (Argentina). We would like to thank Nathan Hageman
for reading and commenting on the manuscript.

\end{acknowledgments}

\newpage

\begin{table}
\caption{Deviations on March 19, 1810}\vskip0.5cm \centering

\begin{tabular}{c c c}
\hline \hline
Time & ~~~~~~~~~~Star Name~~~~~~~~~~ & Measured Deviation\\
[0.5ex] \hline
18:10  & Rigel   &  ~~10$^{\circ}$ 4' 24'', 1\,6 \\ %[1ex]

18:50  & $\alpha$ Orion   &    ~~~~~~~~...25'',\ \ \ 5 \\

20:28  & Castor         &     ~~~~~~~~...24'',\ \ \ 6\\

20:35  & Procyon      &     ~~~~~~~~...24'',\ \ \ 9\\

20:38  & Pollux & ~~~~~~~~...29'',\ \ \ 3\\

22:23  & $\alpha$ Hydra & ~~~~~~~~...22'',\ \ \ 6\\

23:02  & Regulus       & ~~~~~~~~...25'',\ \ \ 2\\

02:19  & Spica & ~~~~~~~~...21'',\ \ \ 4\\

04:30  & $\alpha$ Corona Borealis & ~~~~~~~~...22'',\ \ \ 8\\

04:38  &$\alpha$ Serpens & ~~~~~~~~...22'',\ \ \ 3\\

05:22  & Antares & ~~~~~~~~...22'',\ \ \ 5\\

05:30  & $\zeta$ Ophiuchus & ~~~~~~~~...24'',\ \ \ 0\\ \hline
\end{tabular}
\label{mediciones marzo19}
\end{table}

\begin{table}
\caption{Deviations on March 27, 1810}\vskip0.5cm \centering

\begin{tabular}{c c c}
\hline \hline
Time & ~~~~~~~~~~Star Name~~~~~~~~~~ & Measured Deviation\\
[0.5ex] \hline
18:18  & $\alpha$ Orion   &  10$^{\circ}$ 4' 33'', 28 \\ %[1ex]

19:55  & Castor         &     ~~~~~~...27'', 93\\

20:02  & Procyon      &     ~~~~~~...32'', 31\\

20:06  & Pollux & ~~~~~~...32'', 78\\

21:51  & $\alpha$ Hydra & ~~~~~~...28'', 32\\

00:12  & $\beta$ Leo       & ~~~~~~...30'', 21\\

01:47  & Spica & ~~~~~~~...26'', 29\\

02:39  & Arcturus & ~~~~~~...28'', 05\\

03:58  & $\alpha$ Corona Borealis & ~~~~~~...31'', 39\\

04:49  & Antares & ~~~~~~...28'', 19\\

04:58  & $\zeta$ Ophiuchus & ~~~~~~...29'', 64\\

01:04  & $\gamma$ Virgo & ~~~~~~...27'', 80\\

01:18  & $\delta$ Virgo   & ~~~~~~...27'', 34\\

01:24  & $\epsilon$ Virgo & ~~~~~~...31'', 42\\

23:37  & $\delta$ Leo       & ~~~~~~...34'', 02\\ \hline
\end{tabular}
\label{mediciones marzo27}
\end{table}

\begin{table}
\caption{Deviations on October 8, 1810} \vskip0.5cm \centering

\begin{tabular}{c c c}
\hline \hline
Time & ~~~~~~~~~~Star Name~~~~~~~~~~ & Measured Deviation\\
[0.5ex]\hline

19:26   & $\alpha$ Aquila    & ~22$^{\circ}$ 25' 09''\\

21:24   & Moon crater      & ~~~~...25' 09''\\

21:40   & $\alpha$ Aquarius   & ~~~~...25' 02''\\

02:35   & $\alpha$ Cetus & ~~~~...25' 03''\\

04:08   & Aldebaran & ~~~~...25' 00''\\

04:48   & Rigel & ~~~~...24' 59''\\

05:28   & $\alpha$ Orion & ~~~~...25' 02''\\

06:19   & Sirius & ~~~~...25' 08''\\ \hline
\end{tabular}
\label{mediciones octubre}
\end{table}

\newpage

\newpage

\begin{thebibliography}{99}
\bibitem{foot1}Fizeau also established that the dragging in moving air was not
detectable.
\bibitem{foot2} Since the rays are perpendicular to the wave front in the
ether frame, the starlight aberration should be understood as a
complicated process where the wave fronts change their orientation
when passing from the universal ether to the local terrestrial
ether. A model of this kind was built by Stokes (Ref.
\onlinecite{Stokes}), but Lorentz found it mechanically
inconsistent (Ref. \onlinecite{lorentz}).
\end{thebibliography}
\end{document}